\documentclass[reprint,onecolumn,groupedaddress,showpacs,notitlepage,longbibliography]{revtex4-1}

\usepackage{graphicx}
\usepackage{caption}
\usepackage{subcaption}
\usepackage{color}
\usepackage{amsmath}
\usepackage{amssymb}
\usepackage{verbatim}
\usepackage[english]{babel}
\usepackage{hyperref}

\DeclareGraphicsExtensions{.pdf,.png}

\begin{document}

\title{Sparse model from optimal nonuniform embedding of time series}

\author{Chetan Nichkawde}
    \email{chetan.nichkawde@mq.edu.au}

\affiliation{Department of Physics, Macquarie University, Sydney, Australia}

\date{\today}

\pacs{05.45.Tp, 89.70.Eg, 89.75.-k, 07.05.Tp, 05.45.Pq, 89.65.Gh, 05.10.-a, 05.10.Gg, 05.40.-a}

\begin{abstract}
An approach to obtaining a parsimonious polynomial model from time series is proposed.
An optimal minimal nonuniform time series embedding schema is used to obtain a time delay kernel.
This scheme recursively optimizes an objective functional that eliminates maximum number
of false nearest neighbors between successive state space reconstruction cycles.
A polynomial basis is then constructed from this time delay kernel. 
A sparse model from this polynomial basis is obtained by solving a regularized least squares problem.
The constraint satisfaction problem is made computationally tractable by keeping the ratio between the number of constraints
to the number of variables small by using fewer samples spanning all regions of the reconstructed state space. 
This helps the structure selection process from an exponentially large combinatorial search space.
A forward stagewise algorithm is then used for fast discovery of the optimization path.
Results are presented for the Mackey-Glass system.
\end{abstract}

\maketitle

Modeling and forecasting has been a part of the human quest for knowledge since time immemorial.
At the heart of all time series modeling methods is the notion of causality.
Forecasting relies on the ability to infer the future by observing a system's past.
Modeling involves finding the \emph{rules} that govern the evolution of the
states of the system.
Discovering and quantifying these \emph{rules} lie at the heart of all physical sciences.
Over the past decade a new kind of science\cite{Wolfram} has emerged that seeks to identify
hidden patterns in apparently complex systems.
Complexity is no longer a paranoia with the seminal paper by Watts and Strogatz\cite{Watts} elaborating
how collective dynamics may emerge from a system with a high number of participating entities.
A complex time series has an inherent geometry and Packard et al\cite{Packard} were first to show that a representative geometry of a dynamical
system can be obtained by using a time series of one of its observables.
Takens\cite{Takens} later showed that the geometrical reconstruction methodology proposed by Packard et al\cite{Packard} is actually an embedding of the original
state space of the system. The embedding theorem says that the dynamical attractor obtained using time delayed version of an observable is diffeomorphic to the original state space.

This paper builds a global model for the system from an observed time series on a polynomial basis derived using optimal time delay coordinates.
Consider a time series of observation of a system $x(t)$. Let the dynamics of the system be described by the following model:
\begin{align}
x(t+1) = f\left(x_1,x_2,...,x_m\right)~.
\label{eq:polynomialmodel}
\end{align}
$m$ is the embedding dimension of the system and $x_1=x(t),~x_2=x(t-\tau_1),...,~x_m=x(t-\tau_{m-1})$
are time delay coordinates.
The delays $\tau_1,\tau_2,...,\tau_m$ are determined using a procedure that was shown to be both optimal and minimal\cite{Nichkawde}.
The function $f$ in general is a nonlinear function of its inputs. Polynomials provide good approximation
to $f$ with proper structure selection\cite{letellier2009frequently}. Polynomials have been successfully used to model sun spot time series\cite{aguirre2008forecasting}, Canadian lynx cycles\cite{maquet2007global}, dynamics of mixing reactor\cite{letellier1997recovering}, sleep apnea dynamics\cite{aguirre2004stability} and experimental data from Chua's circuit\cite{aguirre1997nonlinear}.
The model accuracy may be improved by considering polynomials of higher degree. However, parsimony principle
states that the best model is not just that which minimizes error on the observed time series, but should also do
so by having a minimum number of terms. Thus, for the same amount of modeling error on the observed data 
a sparse model with fewer terms is preferred. Overparameterized models lead to overfitting.
The notion of sparsity in parameter space was also shown to be the basis of emergent theories in a recent paper\cite{Machta01112013}.
There are two opposing requirements of increasing accuracy: (1) including polynomial terms of higher degree
and (2) at the same time ensuring sparsity. The number of terms in the polynomial expansion of $f$ grows
with degree $n$ of the polynomial. One must consider $\sum_{i=1}^n\frac{(i+m-1)!}{(i)!(m-1)!}$ terms for a polynomial of degree $n$ with $m$ variables. 
Thus, 20 terms have to be considered for a polynomial of degree 2 with 5 variables,
where as for a polynomial of degree 10 with 5 variables 3002 terms will have to be considered. 
Building a good polynomial model is about structure selection which is a combinatorial optimization problem with search
space exponentially large.
Modeling a system with high embedding dimension imposes great challenge when a polynomial basis
of higher degree is considered, especially when the sample size for modeling is small.

\begin{figure}[t]
\includegraphics[width=0.5\textwidth]{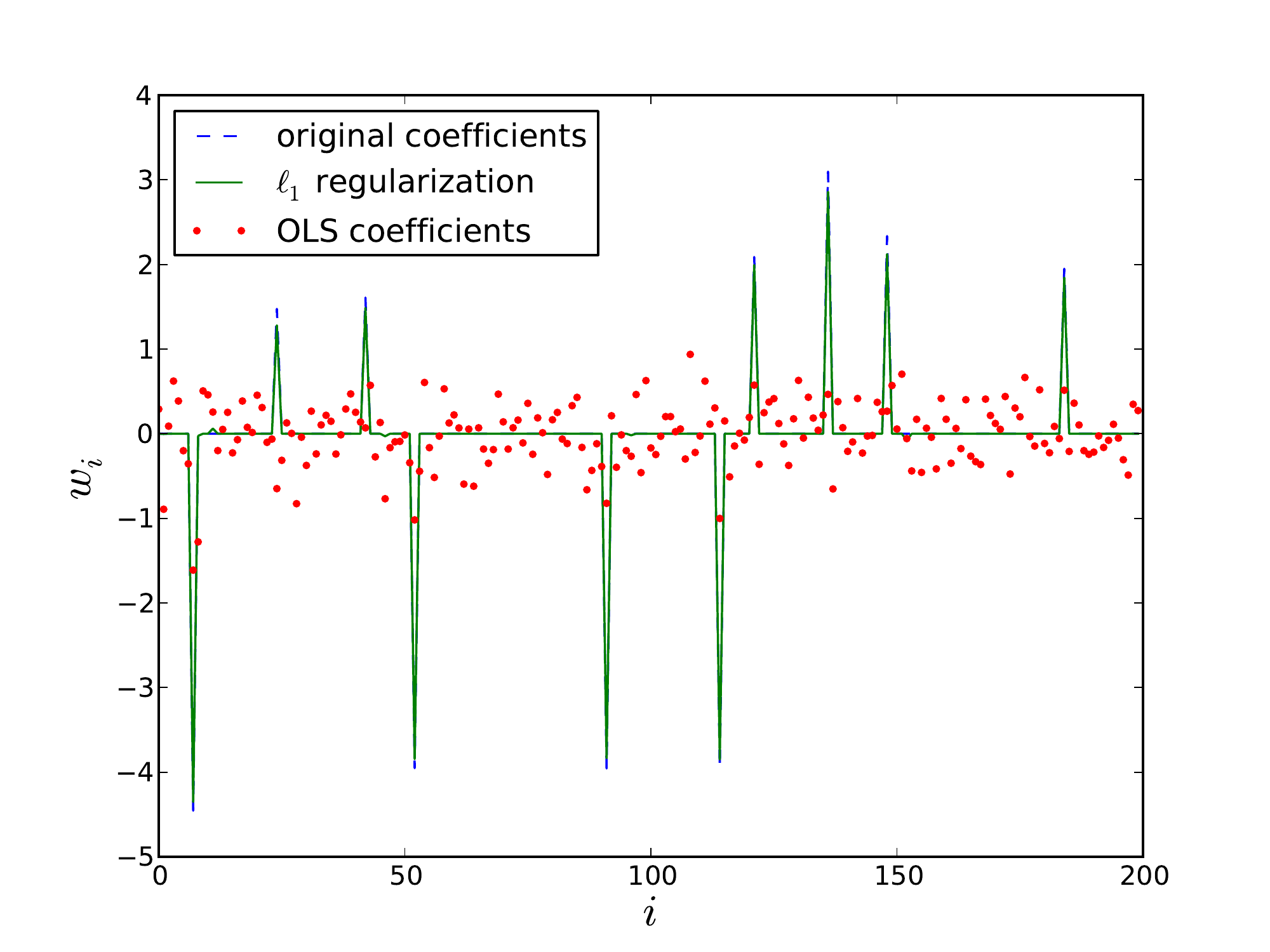}
\caption{\label{fig:sparserecovery} The solution to Eq.~(\ref{eq:linearmodel}) with 200 variables and 50 data points. The regressors $X_i$ are random numbers. 1\% noise was introduced in $y_i$. The solution is sparse with 10 out of 200 $w_i$ being non-zero. Orthogonal least squares (OLS) gives non-zero values to all coefficients. The recovery of coefficients is perfect for $\ell_1$ regularized regression.}
\end{figure}
Consider a function $F$ of $m$ variables $X_1,X_2,...,X_m$.
\begin{align}
y = F(X_1,X_2,...,X_m)~.
\end{align}
$y$ would equal to the future value $x(t+1)$ for an autoregressive function.
The example taken here considers a general function approximation.
Let $F$ be linear and sparse
\begin{align}
y = w_0+w_1X_1+w_2X_2+...+w_mX_m~.
\label{eq:linearmodel}
\end{align}
If $F$ is sparse most of the coefficients $w_1,w_2,...,w_m$ would be zero.
Let us collect $N$ samples of $y$ where $N$ is order $\mathcal{O}(m)$.
The objective is to model $F$ using these observations.
Let $\hat{y}_i$ be the predicted value of the model and let $y_i$ be the true value for the $i^{th}$ data point. $\hat{y}_i$ and $y_i$ would be the one step ahead predicted and true values respectively for an autoregressive model.
\begin{align}
L = \sum_{i=1}^N \left( \hat{y}_i - y_i \right)^2~.
\label{eq:ols}
\end{align}
The error $L$, as given by Eq.~(\ref{eq:ols}), in the modeling ought to minimized.
The usual methodology of linear regression is orthogonal least squares (OLS) which uses the Gram-Schmidt procedure to determine the coefficients $w_i$s such that $L$ is minimized.
As noted previously, the best model is not just the one that minimizes error but must do so with minimal complexity.
OLS has been used in Ref.~\cite{aguirre2008forecasting, maquet2007global, letellier1997recovering, aguirre2004stability, aguirre1997nonlinear} wherein structure selection has been 
done using error reduction ratio criterion. 
Another method for structure selection involves use of Akaike information criterion\cite{barahona1996detection}.
In this procedure, the following information criterion is minimized in accordance with the parsimony principle:
\begin{align}
C(r) = \log \epsilon(r) + \frac{r}{N}~, \nonumber
\end{align}
where $r$ is the number of the chosen term in the polynomial model and is representative of a particular combination of polynomial terms, and,
\begin{align}
\epsilon(r) = \frac{\sum_{i=1}^N \left( \hat{y}_i - y_i \right)^2}{\sum_{i=1}^N \left( y_i - \bar{y} \right)^2}~.
\end{align}
$\bar{y}$ is the mean value of $y_i$ or
\begin{align}
\bar{y} = \frac{1}{N} \sum_{i=1}^N y_i~.
\end{align}
These procedures are obviously time consuming and involve an exhaustive search over exponentially large combinatorial space.

In this paper, learning and structure selection is combined into an efficient single step procedure.
A model complexity term is added to the objective functional in Eq.~(\ref{eq:ols}) that penalizes the $\ell_1$ norm of the model parameters.
Such a form of linear regression is termed as \textit{least absolute shrinkage and selection operator}(LASSO)\cite{tibshirani1996regression} with the objective functional being of the following form
\begin{align}
L_{lasso} = \frac{1}{2N}\sum_{i=1}^N \left( \hat{y}_i - y_i \right)^2 + \alpha \sum_{j=1}^m |w_j|
\label{eq:lasso}
\end{align}
where the hyperparamter $\alpha \geq 0$.
LASSO is known to recover a sparse model\cite{tibshirani1996regression}.
The power of $\ell_1$ based regularization in sparse model recovery is shown in Fig.~\ref{fig:sparserecovery}.
Let $X_1,X_2,...,X_m$ in Eq.~(\ref{eq:linearmodel}) be independent and identically distributed random variables.
Let $m=200$ and only let 10 of the $w_i$s be non-zero. 50 samples of $y$ and corresponding $X_i$s are collected. 
1\% of noise is introduced in $y$.
$\ell_1$ based regularization is able to recover all 10 non-zero coefficients almost perfectly compared to OLS
where all coefficients are erroneously found to be non-zero. 

In this paper, the power of $\ell_1$ based regularization is exploited in sparse model recovery to construct parsimonious polynomial
model for a chaotic system using measured time series. In the next Section, a probabilistic interpretation for the proposed objective function is provided.
In Section~\ref{sec:conssatf}, a constraint satisfaction perspective to the problem is given. In Section~\ref{sec:soln}, the solution procedure is explained. 
Section~\ref{sec:highmackey} explores the model for a high-dimensional version of the Mackey-Glass system. Section~\ref{sec:conclusions} concludes the paper.

\section{\label{sec:probintp}Probabilistic interpretation}
\begin{figure}
\includegraphics[width=0.5\textwidth]{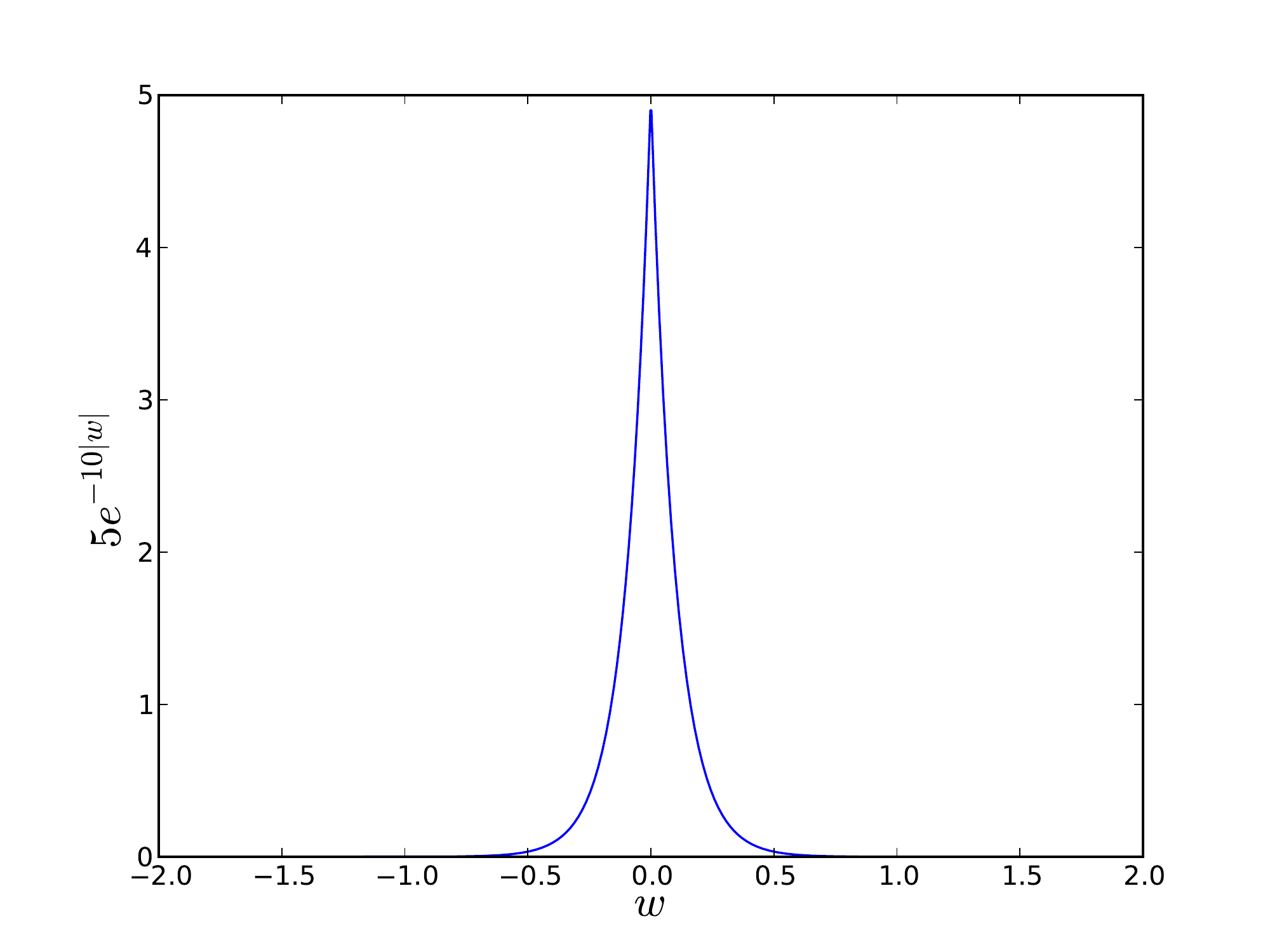}
\caption{\label{fig:laplace} Laplace distribution $\frac{\alpha}{2}e^{-\alpha|w-\mu|}$ with $\alpha=10$ and $\mu=0$.}
\end{figure}

From a Bayesian perspective, the $\ell_1$ term in Eq.~(\ref{eq:lasso})
imposes a Laplacian prior on the parameter space. The Laplace distribution has a sharp probability peak at zero as shown in Fig.~\ref{fig:laplace} for a Laplace distribution with $\alpha=10$. The prior guess about most parameters being zero is thus imposed in this form.
Bayes rule states that the posterior probability of a model is proportional to the product of likelihood and prior
\begin{align}
p(\mathbf{w}|\mathbf{X}) \propto p(\mathbf{X}|\mathbf{w})p(\mathbf{w})~.
\label{eq:posterior}
\end{align}
The $\mathbf{X}$ above is the data, which for our case are values in $m$ dimensional space of regressors $X_j~(j=1,...,m)$.
$\mathbf{w}$ is the vector of coefficients in Eq.~(\ref{eq:linearmodel}).
The residual $\hat{y}_i-y_i$ can be assumed to be Gaussian distributed with zero mean.
The likelihood can thus be defined as
\begin{align}
p(\mathbf{X}|\mathbf{w}) = \left(\frac{1}{N\sqrt{2\pi}}\right)^N\prod_{i=1}^N e^{\frac{-(\hat{y}_i-y_i)^2}{2N}}~.
\end{align}
The Laplacian prior imposed on $\mathbf{w}$ takes the form
\begin{align}
p(\mathbf{w}) = \left(\frac{\alpha}{2}\right)^m\prod_{j=1}^m e^{-\alpha|w_j|}~.
\end{align}
Thus, the negative log of the posterior in Eq.~(\ref{eq:posterior}) can be written as
\begin{align}
-\log \left(p(\mathbf{w}|\mathbf{X})\right) \propto \frac{1}{2N}\sum_{i=1}^N \left(\hat{y}_i-y_i\right)^2 + \alpha \sum_{j=1}^m |w_j|~.
\label{eq:logposterior}
\end{align}
The right hand side of Eq.~(\ref{eq:logposterior}) is exactly equal to the right hand side of Eq.~(\ref{eq:lasso}).
Since logarithm is a monotonic function, the solution to LASSO maximizes the posterior probability of a model with a Laplacian prior.
One possible solution strategy could be to use Markov chain Monte Carlo. Such a random walk is designed to be ergodic on the posterior distribution of $\mathbf{w}$. Due to ergodicity, it draws more samples from near the hills of the landscape, rather than the ravines. Consequently, the samples drawn are representative of the probability distribution. The values of $\alpha$ can be drawn from gamma distribution which is often used in Bayesian inference for width of a distribution. Expectation can be taken on this sample in order to determine $\mathbf{w}$. This approach frees us from making a choice of $\alpha$. However, since this is sampling based, it is a slow solution procedure. 
Moreover, a unique solution is not guaranteed on multiple runs on the same data. A much faster deterministic algorithm is used which will be described later.

\section{\label{sec:conssatf}Constraint satisfaction interpretation}
This can also be seen from a constraint satisfaction problem perspective. $N$ constraints are imposed via Eq.~(\ref{eq:linearmodel}) on $m$ variables $w_1,w_2,...,w_m$.
The problem can be viewed as minimization of $\sum_{j=1}^m |w_j|$ subject to $|y_i-\sum_{j=1}^m w_j X_{ij}| < \epsilon$ where $X_{ij}$ is the value of $j^{th}$ regressor at the $i^{th}$ point
and $\epsilon$ is a small positive number. Computational tractability of constraint satisfaction problems has been studied in various other contexts such as coloring of a random graph\cite{zdeborova2007phase}. 
Notionally, it is harder to satisfy a larger number of constraints simultaneously and there occurs a phase transition in computational landscape beyond which an efficient solution does not exist.
Such problems belong to NP complete class\cite{ercsey2011optimization}.
Thus, the ratio of the number of constraints $N$ and the number of variables $m$, $\frac{N}{m}$, should be kept small so that the optimization task is computationally tractable. 
Large values of $\frac{N}{m}$ result in NP complete optimization problems\cite{mezard2002analytic, ercsey2011optimization}.
This means that an optimal solution can only be found in exponential time or $\mathcal{O}(e^N)$.
The computational landscape analysis performed in the thermodynamic limit of $N$ and $m$ is borrowed in devising an efficient solution procedure.

\section{\label{sec:soln}Solution}
\subsection{Optimal Minimal Embedding}
\begin{figure*}
    \centering
    \begin{subfigure}[t]{0.45\textwidth}
        \centering
        \includegraphics[width=\textwidth]{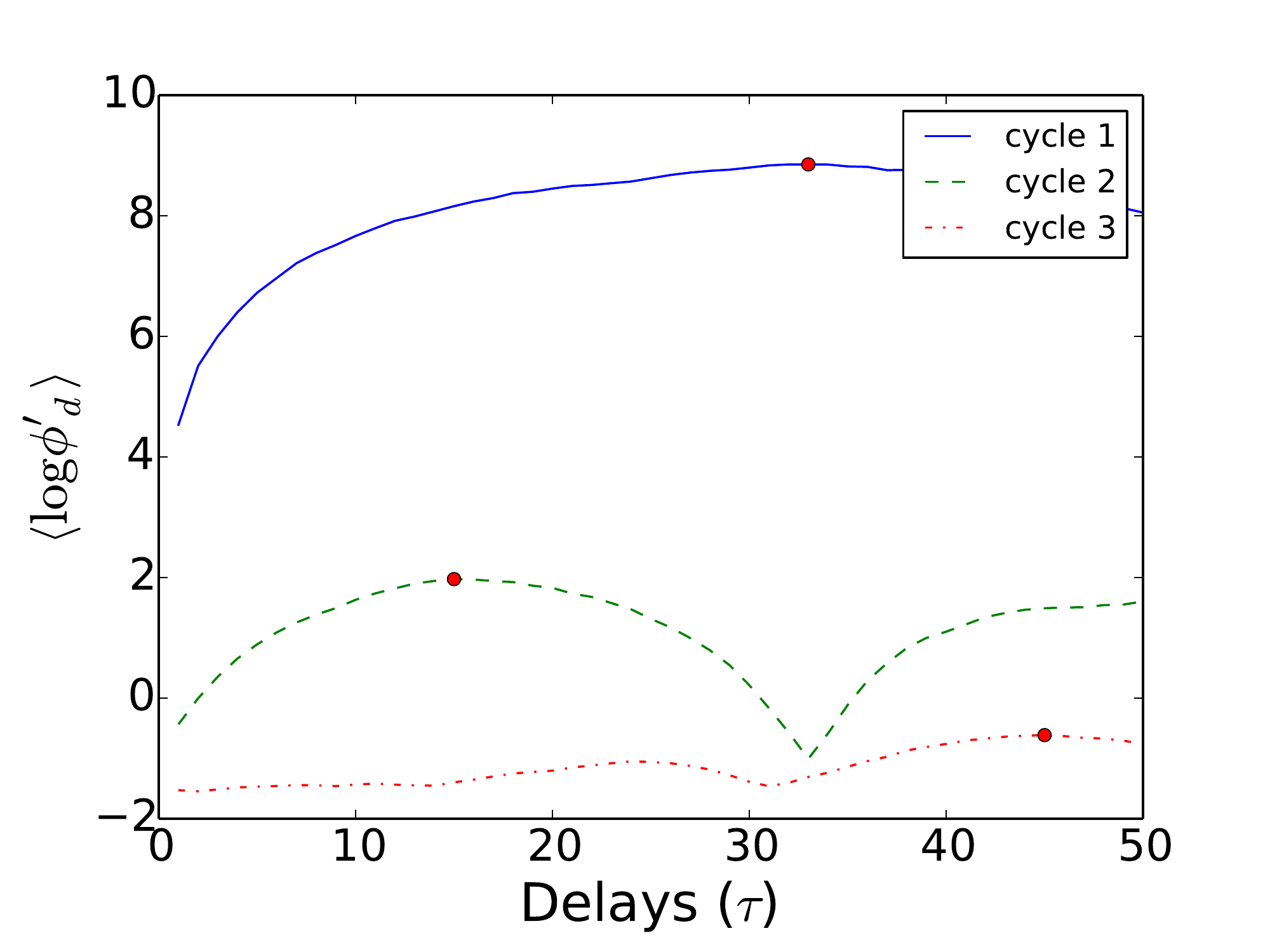}
        \caption{Reconstructions cycles for Mackey-Glass system showing most optimal delays from 3 cycles of recursive optimization as 33, 15 and 45. Parameter values are given in the text.}
        \label{fig:mackey23_recr}
    \end{subfigure}
    \hfill
    \begin{subfigure}[t]{0.45\textwidth}
        \centering
        \includegraphics[width=\textwidth]{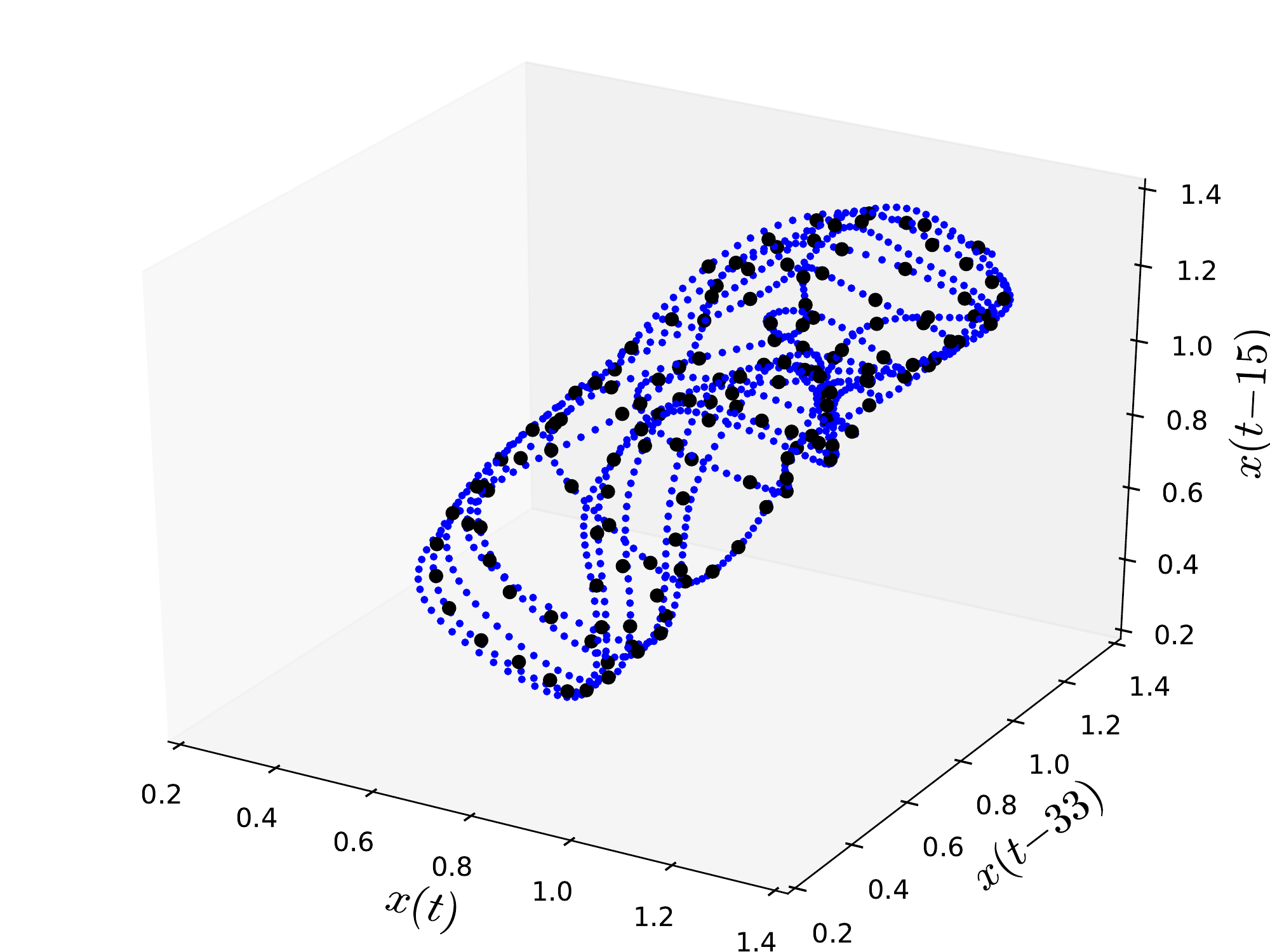}
        \caption{Clusters and their centroid for reconstructed Mackey-Glass attractor with delays 33, 15 and 45 with 1000 points.}
        \label{fig:mackey23_2000_cluster}
    \end{subfigure}
    \caption{State space reconstruction and reconstructed attractor showing cluster centroids.}
\end{figure*}
\setcounter{subfigure}{0}
Takens embedding theorem allows us to reconstruct the representative state space of a dynamical system\cite{Takens}.
The hallmark of this result is that a state space can be reconstructed using only one or a few of the observables.
The first time delay is chosen as the first minimum of mutual information between observed variable $x(t)$ and its time-delayed
counterpart $x(t-\tau)$ in order to determine the optimal time delay $\tau$\cite{Mutual}.
This function is given by
\begin{align}
I(x_1,x_2) = \sum p(x_1,x_2) \log \left(\frac{p(x_1,x_2)}{p(x_1)p(x_2)}\right)
\end{align}
where $x_1=x(t)$ and $x_2=x(t-\tau)$.
The subsequent delays are chosen in an \emph{ad hoc} manner as multiples of $\tau$ ($2\tau,~3\tau,...,(m-1)\tau$).
This approach is termed as uniform as the chosen time lags are uniformly separated.
The uniform time delay approach is however not always optimal for modeling\cite{aguirre2009modeling}.
Investigations have also shown that nonuniform embedding performs better than uniform embedding for causality and coupling detection\cite{Vlachos, Luca, kugiumtzis2013direct}.
The embedding dimension $m$ above is determined by false nearest neighbors approach\cite{Kennel}.
In this approach, a nearest neighbor is flagged as false, if the same two points become distant in a higher embedding dimension.
$m$ is thus chosen when there are no false nearest neighbors. 
Recent investigations have revealed two major caveats with this approach\cite{Nichkawde}. 
\begin{enumerate}
\item That the inability to eliminate false nearest neighbors could be due to a poor choice of the last delay $(m-1)\tau$. Please refer to Section III in Ref.~\cite{Nichkawde}.
\item The uniform delay approach is not a minimal approach. It is possible to eliminate larger number of false nearest neighbors between successive reconstruction cycles and thus achieve an embedding with a smaller number of delays.
Minimality is important in Occam's razor principle. 
The minimum description length principle\cite{rissanen1978modeling} also stresses minimality by having size or complexity of model a part of the description length.
\end{enumerate}
Another very important concept is irrelevancy\cite{Casdagli}. The last few longer delays in the uniform embedding approach may be irrelevant. Causality is lost beyond a certain time in the past.

Addressing the above highlighted issues, an optimal nonuniform embedding of the system is used for modeling\cite{Nichkawde}.
The methodology imposes an upper bound in the search for optimal time delays by using an objective functional that quantifies irrelevancy.
The reconstruction methodology is very fast and performs in $\mathcal{O}(N \log N)$ where $N$ is the length of the time series.
This methodology recursively chooses delays that maximize derivatives on the project manifold.
The objective functional is of the following form:
\begin{align}
\log \left[\beta_d(\tau_d)\right] = \left<\log \phi'_{d_{ij}}\right>~.
\label{eq:mdop}
\end{align} 
In the above equation, $\phi'_{d_{ij}}$ is the absolute value of the directional derivative evaluated in the direction from the $i^{th}$ to the $j^{th}$ point of the projected attractor manifold
which happens to be the nearest neighbor and $\left<\cdot\right>$ stands for the mean value.
$\phi'_{d_{ij}}$ may also be expressed as follows
\begin{align}
{\phi_d'}_{ij}^2(\tau_d) = \frac{R^2_{d+1}(i)-R^2_d(i)}{R^2_d(i)}
\label{eq:phird}
\end{align} 
where $R_d(i)$ be the Euclidean distance of the the $i^{th}$ point in the ${(d-1)}^{th}$ reconstruction cycle to its nearest neighbor and $R_{d+1}(i)$ is the same distance in the $d^{th}$ reconstruction cycle with $x(t-\tau_d)$ as the additional time delay coordinate.
The nearest neighbor in the ${d-1}^{th}$ cycle is deemed as false if:
\begin{align}
\frac{R^2_{d+1}(i)-R^2_d(i)}{R^2_d(i)} > R_{tol}
\end{align}
where $R_{tol}$ is a threshold\cite{Kennel}.
Now using Eq.~(\ref{eq:phird}) above can be written as:
\begin{align}
{\phi_d'}_{ij}^2(\tau_d) > R_{tol}~.
\label{eq:phirtol}
\end{align}
The right hand side of Eq.~(\ref{eq:mdop}) is the geometric mean of ${\phi_d'}_{ij}$. It is thus obvious from the above relationship that the recursive maximization of the functional given by Eq.~(\ref{eq:mdop}) eliminates the largest number of false nearest neighbors between successive reconstruction cycles and thus helps achieve an optimal minimal embedding\cite{Nichkawde}.

Fig.~3 shows the log-likelihood of derivatives on projected manifold given by Eq.~(\ref{eq:mdop}) for the Mackey-Glass system\cite{mackey1977oscillation}.
Mackey-Glass equation is the nonlinear time delay differential equation
\begin{align}
\frac{dx}{dt} = \beta \frac{x_\tau}{1+x_\tau^n}-\gamma x, \quad \gamma,\beta,n > 0 
\label{eq:emackey}
\end{align}
where $x_\tau$ represents the value of $x$ at time $t-\tau$.
The equation shows chaos with increasing complexity with an increasing value of $\tau$.
The time series was generated with parameter values $\gamma=0.1,~\beta=0.2,~n=10$ and $\tau=23$.
2000 points were sampled with $\delta t$ of 0.5 after removing the transient.
First 1000 points are used for model building.
Delays of 33, 15 and 45 shown by round dots are found to be most optimal as shown in Fig.~\ref{fig:mackey23_recr}.

\subsection{Polynomial model from optimal nonuniform delays} 
Continuing to develop a polynomial model for the Mackey-Glass example, a polynomial model of 4 variables $x_1=x(t),~x_2=x(t-33),~x_3=x(t-15)$ and  $x_4=x(t-45)$ in Eq.~(\ref{eq:polynomialmodel}) is considered.
$X_i$s in Eq.~(\ref{eq:linearmodel}) are replaced by a polynomial basis of order $n$ of variables $x_1,~x_2,~x_3$ and $x_4$.
\begin{align}
x(t+1) & =  w_0 + w_1 x_1 + w_2 x_2 + w_3 x_3 + w_4 x_4 \nonumber \\
  &    + w_5 x_1^2 + w_6 x_2^2 + w_7 x_3^2 + w_8 x_4^2 \nonumber \\
  &    + w_9 x_1 x_2 + w_{10} x_1 x_3 + w_{11} x_1 x_4 + w_{12} x_2 x_3 \nonumber \\
  &    + w_{13} x_2 x_4 + w_{14} x_3 x_4 + \, ....
\label{eq:polyterm}
\end{align}
Terms up to degree 2 have been shown in the equation above.
Let there be $m$ model terms with corresponding coefficients $w_1,w_2,...,w_m$. Let us collect $N$ measurements and plug those in Eq.~(\ref{eq:polyterm}).
This system of $N$ equations is overdetermined if $N > m$. 
Since there are $m$ unknowns, only about $m$ number of measurements is required.
This fact is used and an effective sample of $\mathcal{O}(m)$ number of measurements is drawn to solve the inverse problem. 
This is achieved by using $k$-means clustering algorithm\cite{Lloyd} which samples points approximately from all regions of the reconstructed attractor.
Given a set of $N$ observations $\mathbf{x_j}$, $j=1,2,...,N$, $k$-means clustering aims to partition the $N$ observations into $k$ sets $\mathbf{S} = \{S_1, S_2, ..., S_k\}$ so that within each cluster, the Euclidean distance
of the points from the cluster mean $\mathbf{\mu_i}$ of $S_i$
\begin{align}
\underset{\mathbf{S}} {\operatorname{arg\,min}}  \sum_{i=1}^{k} \sum_{\mathbf{x_j} \in S_i} \left\| \mathbf{x_j} - \boldsymbol\mu_i \right\|^2 \nonumber
\end{align}
is minimized.
There are heuristic algorithms which achieve nearly the same result in much faster time\cite{sculley2010web}.
This accomplishes two simultaneous objectives: 1) using only few representative samples greatly speeds up the computation. 2) Fewer samples imply fewer constraints
and therefore this renders the problem more computationally tractable. 
As noted previously in Section~\ref{sec:conssatf}, from a constraint satisfaction perspective large values in $\frac{N}{m}$ lead to a hard optimization problem.
This means that the only possible solution that will work is a sequential search through exponentially large combinatorial search space.
For example, a $7^{th}$ degree polynomial model of Mackey-Glass system with 4 variables would have 329 terms and approximately $1.093625 \times 10^{99}$ possible structures.
Appropriately sampling the reconstructed state space using k-means clustering keeps the ratio $\frac{N}{m}$ small and brings down the computational complexity to polynomial time or $\mathcal{O}(N^d)$.
In this paper, the parameters which minimizes the objective function given by Eq.~(\ref{eq:lasso}) are learnt by least angle regression\cite{Efron} algorithm which discovers the optimization path in a forward stagewise manner.
In this approach, the origin of the residual space (also elsewhere referred to as \textit{data space}\cite{transtrum2010nonlinear,transtrum2011geometry} with manifold formed by various values of parameters as \textit{model manifold}) is approached by including the regressors one at a time which makes the least angle with the current residue.
The optimal value for the parameter $\alpha$ is determined using 5-fold cross validation.
The least angle regression algorithm achieves this in linear time or $\mathcal{O}(N)$.
The values are rescaled between 0 and 1 before building the model.
The proposed approach will work even for small time series as long as the entire range of state space has been sampled.
\begin{figure*}
    \centering
    \begin{subfigure}[t]{0.45\textwidth}
        \centering
        \includegraphics[width=\textwidth]{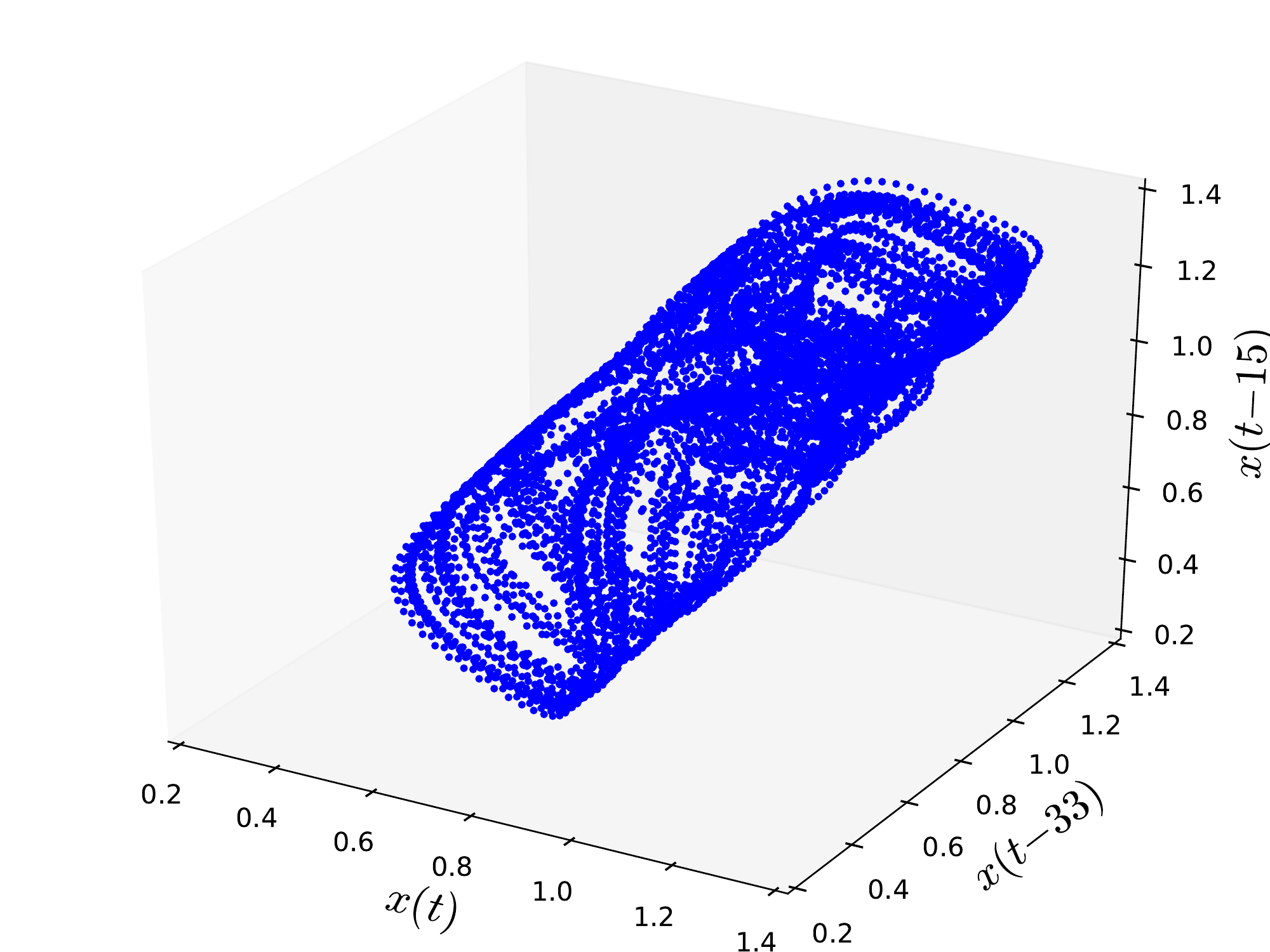}
        \caption{Mackey-Glass attractor using signal generated from the learnt $7^{th}$ degree polynomial model given by Eq.~(\ref{eq:mackey23_2000_order7model}).}
        \label{fig:mackey23_2000_genattractor}
    \end{subfigure}
    \hfill
    \begin{subfigure}[t]{0.45\textwidth}
        \centering
        \includegraphics[width=\textwidth]{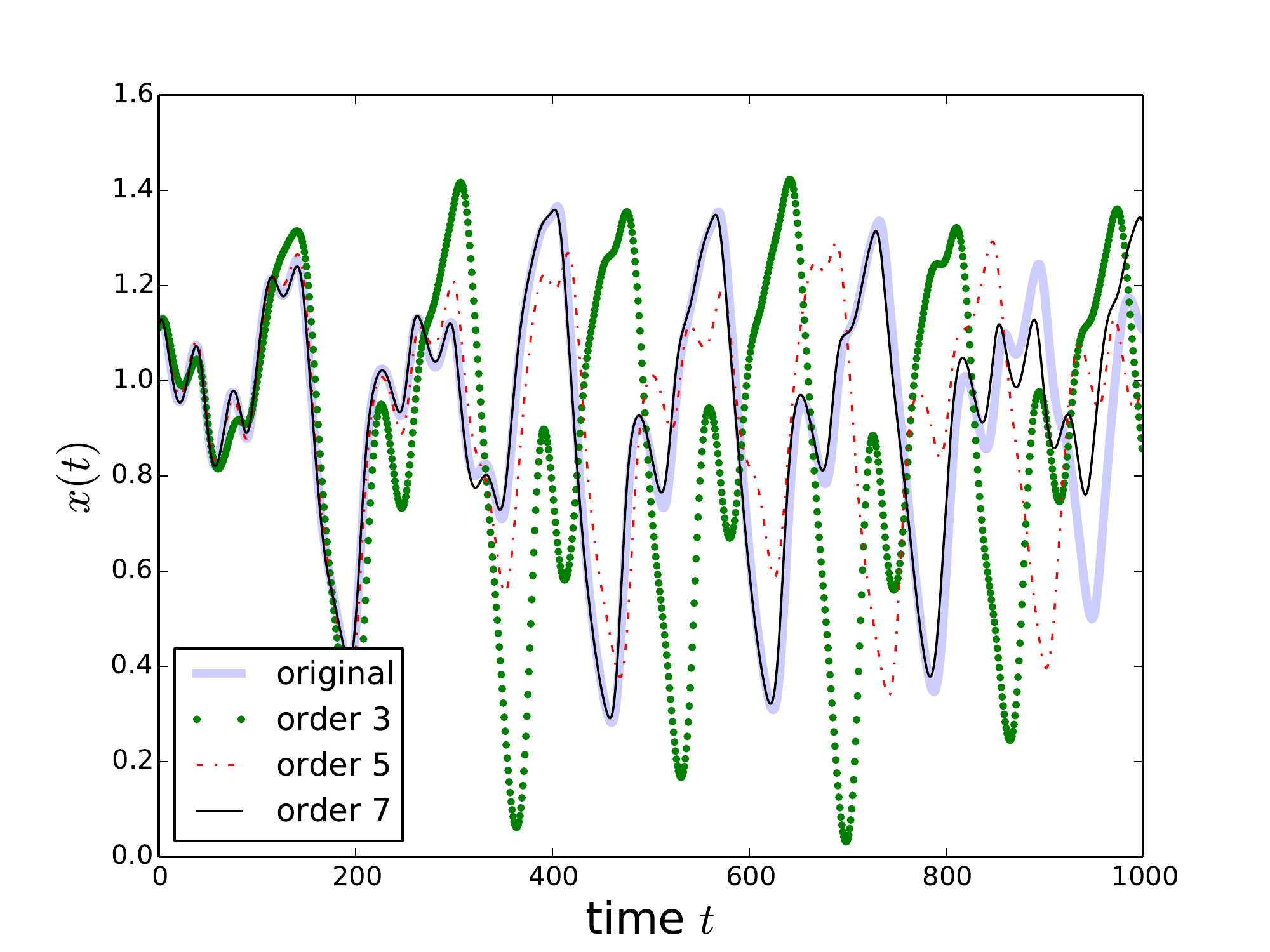}
        \caption{Prediction of the models with polynomial degree 3, 5 and 7. The prediction accuracy improve with degree of the polynomial.}
        \label{fig:mackey23_2000_prediction}
    \end{subfigure} 
    \begin{subfigure}[t]{0.5\textwidth}
        \centering
        \includegraphics[width=\textwidth]{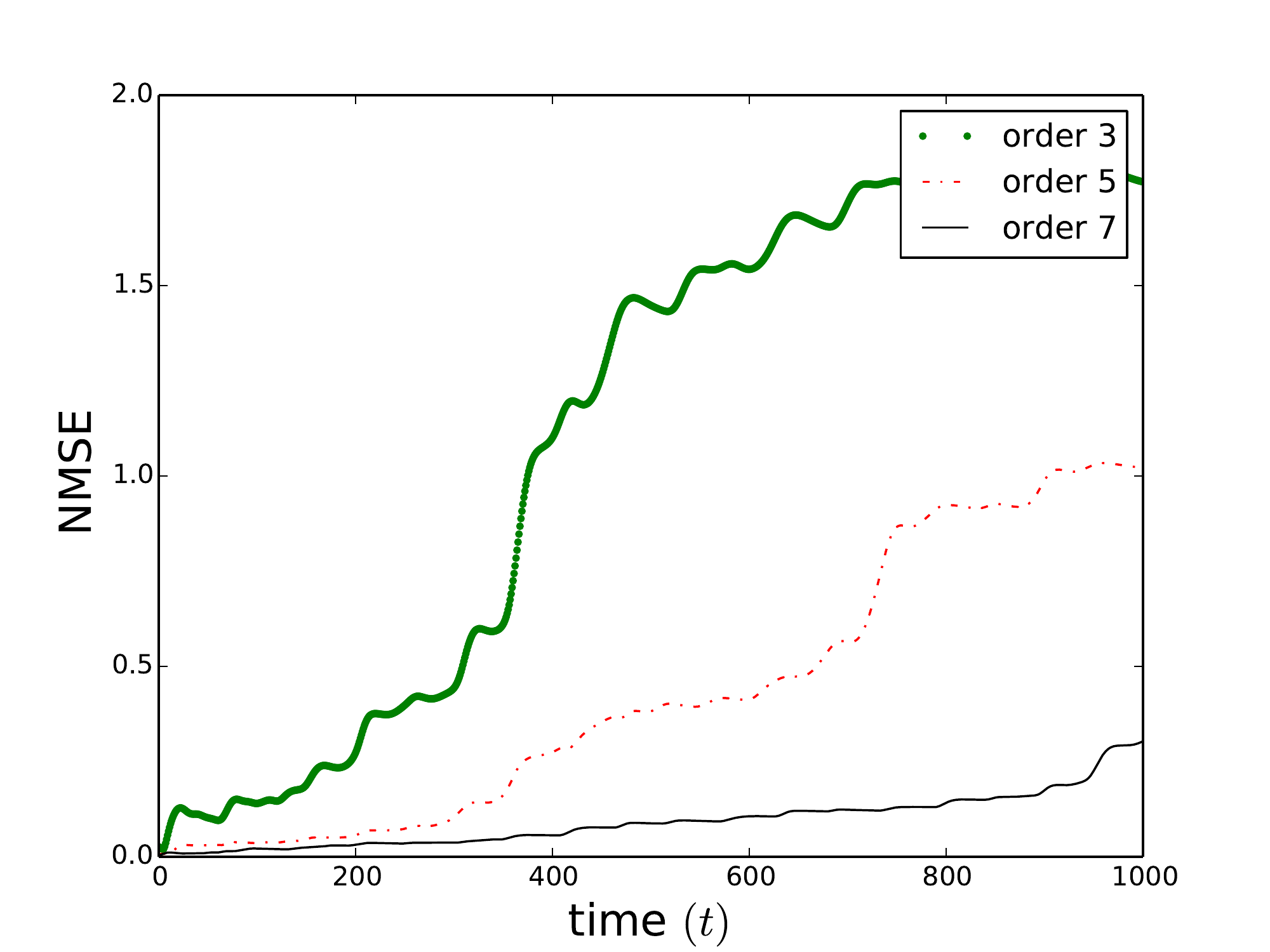}
        \caption{Mean square error with time for data in Fig.~\ref{fig:mackey23_2000_prediction}.}
        \label{fig:mackey23_2000_nmse}
    \end{subfigure}
    \caption{Generated attractor, prediction and mean square error for Mackey-Glass system.}
\end{figure*}
\setcounter{subfigure}{0}
This is demonstrated with the Mackey-Glass time series. 1000 data points are used. These points, however, span almost the entire space of the attractor as shown in Fig.~\ref{fig:mackey23_2000_cluster}.
The clusters for the reconstructed Mackey-Glass attractor and their centroids are also shown in Fig.~\ref{fig:mackey23_2000_cluster}.
The model building is done by randomly sampling a point from each of the clusters.
This model was then used to predict the next 1000 values.

The attractor from the signal generated using the learnt model is shown in Fig.~\ref{fig:mackey23_2000_genattractor}.
Fig.~\ref{fig:mackey23_2000_prediction} shows the predictions for the next 1000 values for models of the Mackey-Glass system of with polynomial degrees 3, 5 and 7. 
The thick solid lines shows the actual values. The model accuracy increases with increased polynomial degree as shown in Fig.~\ref{fig:mackey23_2000_nmse} which shows the normalized mean square error (NMSE)
for models of increasing polynomial degree. Model with polynomial degree 7 has the best NMSE which was shown to be 0.07 after 1000 time steps.
This model has 64 nonzero terms out of a total of 329 terms.

\begin{figure*}
    \centering
    \begin{subfigure}[t]{0.45\textwidth}
        \centering
        \includegraphics[width=\textwidth]{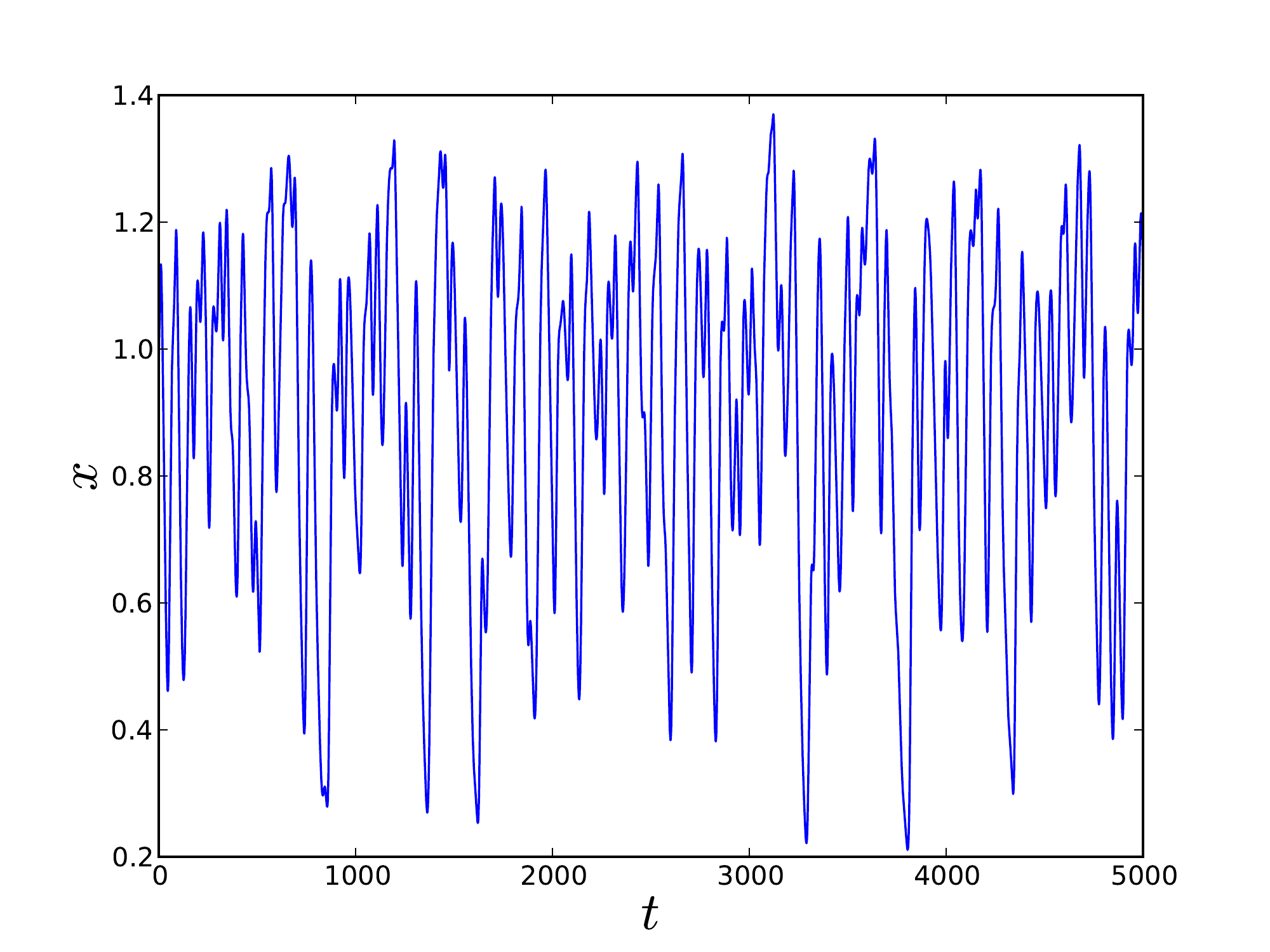}
        \caption{Time series for simulation of high-dimensional Mackey-Glass equation.}
        \label{fig:mackey80series}
    \end{subfigure}
    \hfill
    \begin{subfigure}[t]{0.45\textwidth}
        \centering
        \includegraphics[width=\textwidth]{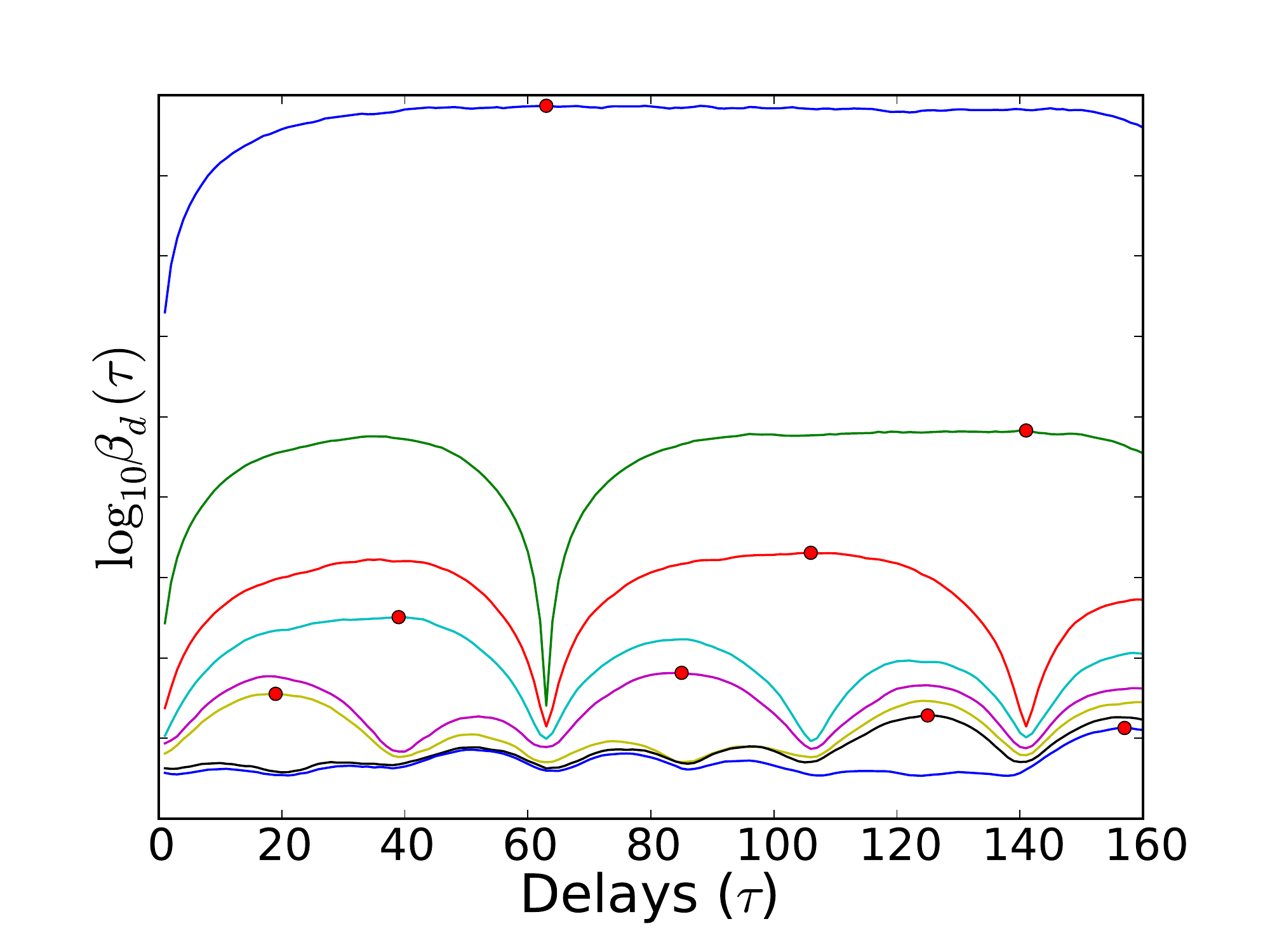}
        \caption{Reconstructions cycles for high-dimensional Mackey-Glass system. Delays of 63, 141, 106, 39, 85, 19, 125 and 157 were found to be most optimal.}
        \label{fig:mackey80_recr}
    \end{subfigure}
    \caption{State space reconstruction of high-dimensional Mackey-Glass system.}
\end{figure*}
\setcounter{subfigure}{0}

\section{\label{sec:highmackey}High-dimensional Mackey-Glass system}
A high-dimensional version of Mackey-Glass model is probed. The delay parameter $\tau$ in Eq.~(\ref{eq:emackey}) is set to 80 and 5000 points are sampled at step size of 0.5.
The time series is shown in Fig.~\ref{fig:mackey80series}.
The Kaplan-Yorke dimension $D_{KY}$ for this system is about 8\cite{poon2001titration}.
State space reconstruction is performed for this time series. The reconstruction cycles have been shown in Fig.~\ref{fig:mackey80_recr}.
Delays of 63, 141, 106, 39, 85, 19, 125 and 157 were obtained with an embedding dimension of 9.
A $7^{th}$ degree polynomial model is built for this series. The model obtained with their terms has been given in Appendix~\ref{sec:mackey80model}.
The model has a total 89 terms selected out of 11439 terms. 
\subsection{\label{sec:modeleval}Model evaluation using anticipated synchronization}
\begin{figure*}
    \centering
    \begin{subfigure}[t]{0.45\textwidth}
        \centering
        \includegraphics[width=\textwidth]{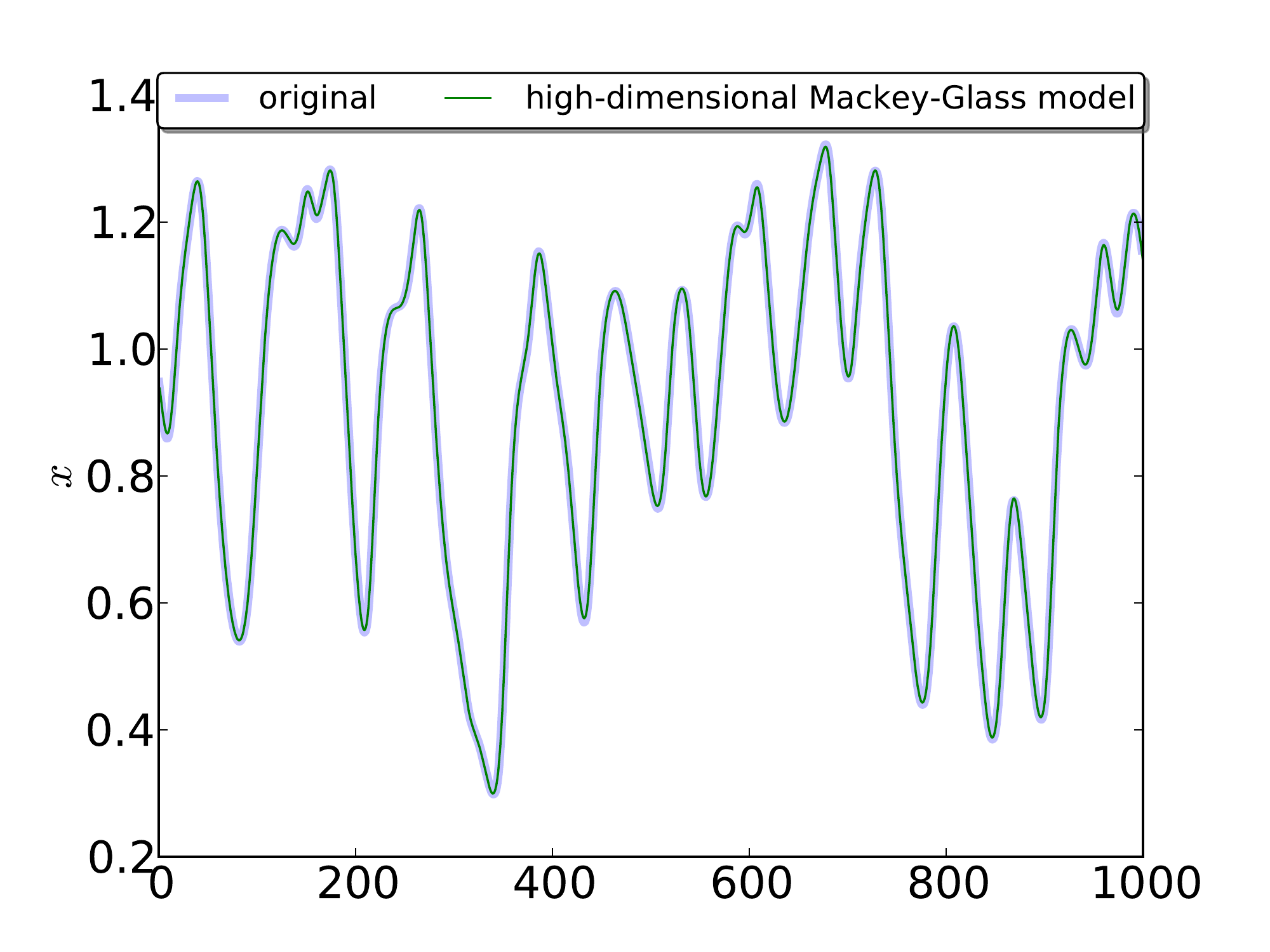}
        \caption{Evolution of high-dimensional Mackey-Glass model given by Eq.~(\ref{eq:mackey80model}) coupled to Mackey-Glass time series. Since, model given by Eq.~(\ref{eq:mackey80model}) is matched to Mackey-Glass time series, it synchronizes with the external driver.}
        \label{fig:mackey80sync}
    \end{subfigure}
    \hfill
    \begin{subfigure}[t]{0.45\textwidth}
        \centering
        \includegraphics[width=\textwidth]{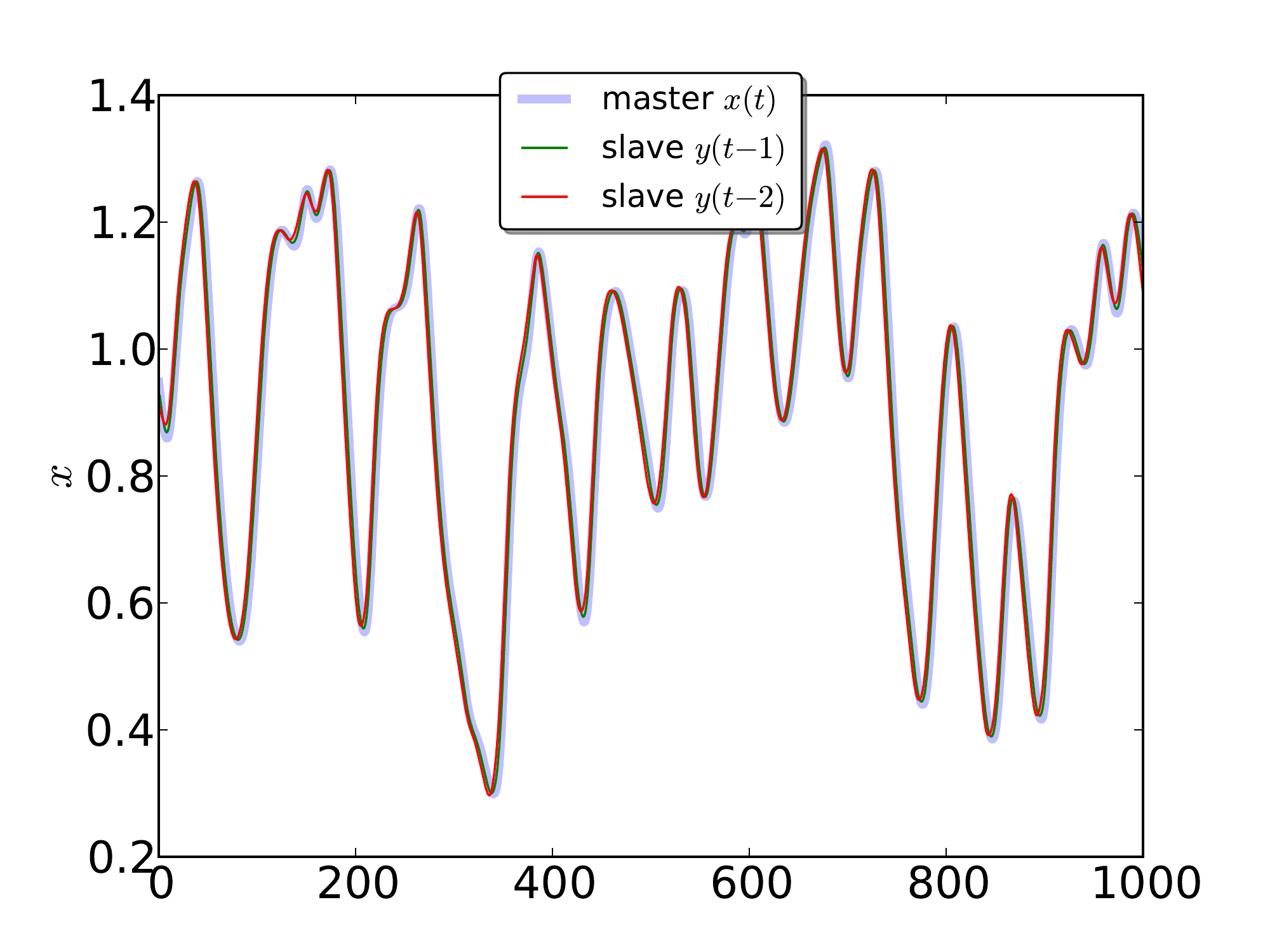}
        \caption{Anticipated synchronization of high-dimensional Mackey-Glass model given by Eq.~(\ref{eq:mackey80model}) coupled to Mackey-Glass time series. Future states can be seen to be anticipated 2 step in advance.}
        \label{fig:mackey80syncanticipated}
    \end{subfigure}
    \begin{subfigure}[t]{0.9\textwidth}
        \centering
        \includegraphics[width=\textwidth,height=0.4\textwidth]{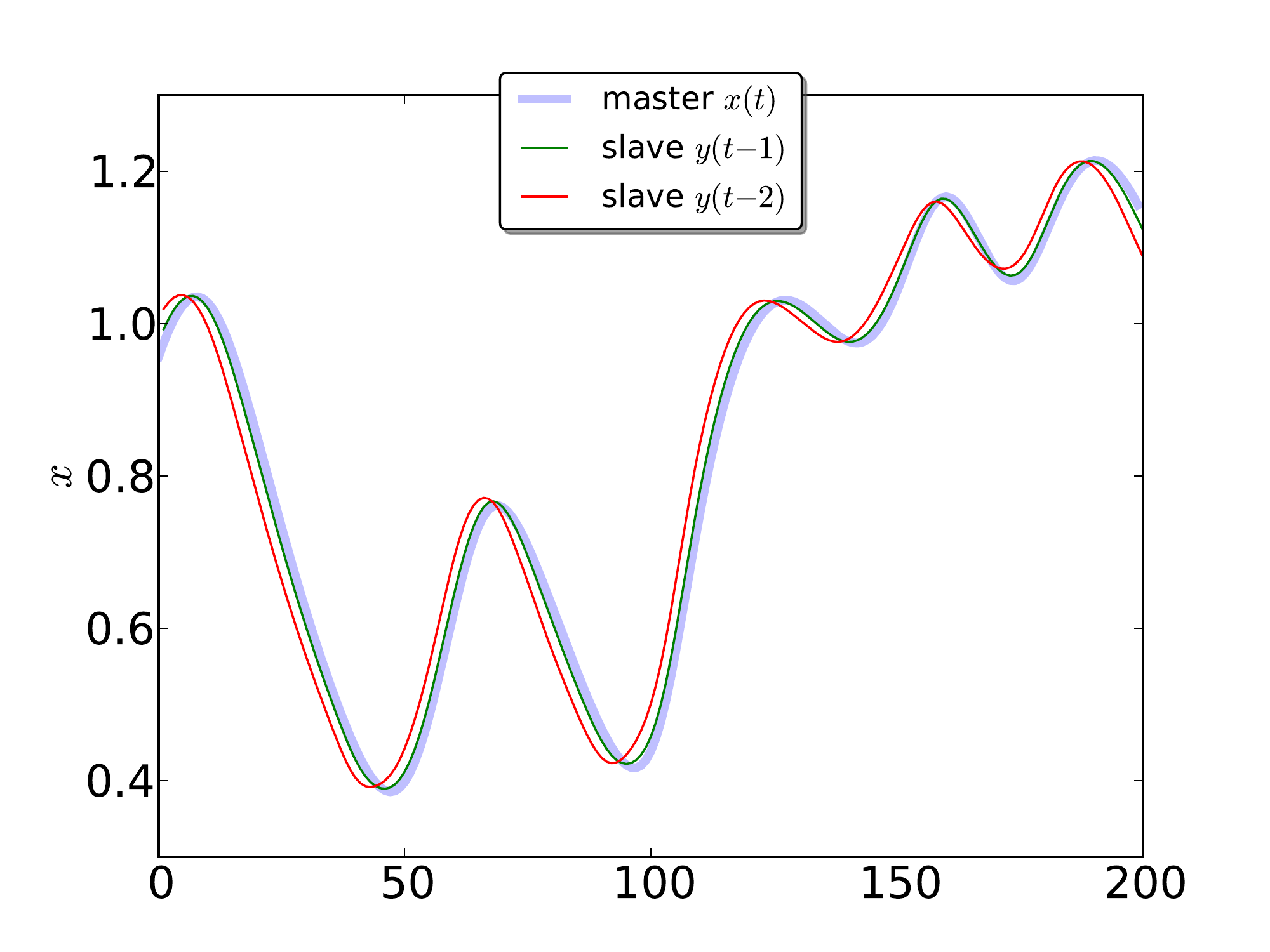}
        \caption{Zooming on last 200 values in Fig.~\ref{fig:mackey80syncanticipated}.}
        \label{fig:mackey80syncanticipatedzoom}
    \end{subfigure}
    \caption{Synchronization of Mackey-Glass time series with the learnt model.}
\end{figure*}

Evaluation of a dynamical model can be done in several different ways. Chaotic attractors have certain geometrical properties such as fractal dimension and Lyapunov exponent which can be compared with that of the original system. 
In cases, where the available time series is very short and high-dimensional, it is not possible to determine these properties from the time series. A different approach is used in this case in order to evaluate the model. An interesting property of matched systems is that they can easily synchronize\cite{brown1994modeling}. This property can be used to evaluate the learnt dynamical model from time series.
Let $\mathbf{G}$ be the true dynamics of the system:
\begin{align}
x(t+1) = \mathbf{G(x(t))}~.
\label{eq:truemodel}
\end{align}
Let $\mathbf{F}$ denote the dynamics of the learnt model.
The learnt model can be coupled to the true model as follows
\begin{align}
y(t+1) = \mathbf{F(y(t))}-k(y(t)-x(t))~.
\label{eq:modelsync}
\end{align}
If $k$ is adequately chosen and $\mathbf{F}$ is sufficiently close to $\mathbf{G}$ then $y_t \to x_t$.
This means that the model will synchronize with the observed values.
If $\mathbf{G}$ and $\mathbf{F}$ differ, the error $e(t)=y(t)-x(t)$ will not go to
zero but will stay around the origin of the residual space. 
The average distance to the origin of such a space will depend on the difference between $\mathbf{G}$ and $\mathbf{F}$.
Therefore such a distance is a measure of how far the estimated model $\mathbf{F}$ is from the true dynamics $\mathbf{G}$\cite{aguirre2006evaluation}.
In our case, the true dynamics $\mathbf{G}$ is unknown. However, the scalar measurement $x(t)$ is available.
$\mathbf{y(t)}$ is taken as an appropriate embedding of the scalar observation and $y(t+1)$ is the prediction of the model. 
A coupling of form $k(y(t)-x(t))$ where $k=0.7$ is used.
The model learnt for this system is synchronized with another set of observations for the same system. These observations were not used in building the model.
Evolution of high-dimensional Mackey-Glass model given by Eq.~(\ref{eq:mackey80model}) coupled to the second set of observations is shown in Fig.~\ref{fig:mackey80sync}.
Since, the model give by Eq.~(\ref{eq:mackey80model}) is matched to the time series dynamics, it synchronizes with the external driver.

A particular form of synchronization known as anticipated synchronization can be used to anticipate the future states in advance\cite{voss2001dynamic}.
In this form of synchronization, a time delay $\tau > 0$ is introduced in Eq.~(\ref{eq:modelsync}).
This introduced memory gives the system
\begin{align}
y(t+1) = \mathbf{F(y(t))}-k(y(t-\tau)-x(t))~.
\end{align}
The driving system is unaffected and only the past values of driven system are needed.
A matched system would synchronize on the manifold given by
\begin{align}
y(t-\tau)=x(t)~,
\end{align}
on which the driven system $y$ anticipates driving system's state $x$. 
This sort of scheme can also be chained where the output of the first slave in the chain $y$ can be made to synchronize with a delayed version of the same model
\begin{align}
z(t+1) = \mathbf{F(z(t))}-k(z(t-\tau)-y(t))~.
\end{align}
Figure~\ref{fig:mackey80syncanticipated} shows the anticipation with $\tau=1$ for two slaves in the chain.
Figure~\ref{fig:mackey80syncanticipatedzoom} zooms onto the last 200 values in Fig.~\ref{fig:mackey80syncanticipated}.
Slave 1 anticipates 1 time step ahead and slave 2 anticipates 2 time steps ahead.
Future states can easily be anticipated in advance using this synchronization scheme.
This demonstrates building of a sparse data-driven model for a very high dimensional system.

\section{\label{sec:conclusions}Discussion and Conclusions}
A new method for developing a polynomial model from complex time series data has been described. 
A generalized linear model setting has been used with polynomial terms obtained from optimal
nonuniform delays as regressors. Penalization of $\ell_1$ norm of model coefficients
leads to recovery of a parsimonious model.
The approach combines structure selection and model fitting into a single step procedure
by discovering the optimization path in a forward stagewise manner.
The solution procedure is guided in terms of both speed and constraint satisfiability by using
fewer samples taken from all regions of reconstructed state space. 
The procedure has been demonstrated on the Mackey-Glass system.
A highly accurate model of a low dimensional Mackey-Glass system was obtained which is able to do long term prediction.
The model for high-dimensional Mackey-Glass system synchronizes very well with the simulated observations.
It is interesting to note that the inverse problem for high-dimensional Mackey-Glass system is underdetermined when a $7^{th}$ degree polynomial model is considered with an embedding dimension of 9. 
There are 11439 variables whereas the number of equations is less than this number for the size of the data. 
The present approach is still able to recover a sparse model. This is because the signal has a sparse representation in a polynomial basis.
Thus, for this case, the solution exactly follows the fundamentals of compressed sensing\cite{candes2008introduction} which has emerged as the state of the art technique in signal acquisition. Measurement matrix is chosen to be comprised of random numbers in compressed sensing\cite{candes2008introduction} whereas the `measurement matrix' $\mathbf{X}$ in this case happens to be values of polynomial regressors.
The constraint satisfaction perspective given in Section~\ref{sec:conssatf} is valid only for an overdetermined system ($N > m$). 
However, when the system is underdetermined ($N < m$) such as the case for high-dimensional Mackey-Glass system, all the $N$ available data points should be used in building the model.
The proposed approach can also be used to detect nonlinear dynamics in time series with the noise titration approach\cite{poon2001titration}. 
In this approach, nonlinear dynamics is ascertained by presence of nonlinear terms in an autoregressive model. 
If a linear model performs as well as a nonlinear model, then it is concluded that nonlinear dynamics is not present in the time series. 
The performance is measured by mean square error over the entire data. As noted previously, it is always possible to reduce such a metric almost indefinitely by overparametrizing. 
Thus, the robust framework proposed in this paper to build a parsimonious model with small size data would be a great choice for the noise titration approach. 
Extending this argument, this approach can also be used to detect the presence of colored nonlinearly correlated noise\cite{freitas2009failure}.
The dynamics in economics and financial settings are often driven by additive white noise. 
This noise component is deemed to be conditionally heteroskedastic. This means that they have varying variance whose value depends on few of the past values. 
This form of modeling is termed as autoregressive conditional heteroskedasticity. The framework proposed in this paper allows us to explore possibly nonlinear conditional heteroskedasticity with perforations.
A polynomial approach to nonlinear Granger causality\cite{granger1969investigating} directly follows from this work. 
One just needs to check if the predictive power of the model is enhanced by including polynomial terms from the second variable. 
The methodology can thus detect the causal effect of linear and nonlinear terms composed of the embedding of the second variable. 

\appendix
\section{\label{sec:mackeymodel}Model for Mackey-Glass system}
\begin{widetext}
The $7^{th}$ degree polynomial model for Mackey-Glass system with parameters value $\gamma=0.1,~\beta=0.2,~n=10$ and $\tau=23$:
\begin{align}
x(t+1) = & 0.383613+1.212485x_1+0.324682x_2+0.293563x_3+0.367696x_4+0.476091x_1x_3 \nonumber \\
&+0.534483x_1x_4+0.353810x_2x_3+0.330629x_2x_4+0.284664x_3^2+0.358371x_1^3 \nonumber \\
&+0.332392x_1^2x_2+0.127815x_1^2x_4+0.302744x_1x_3^2+0.609589x_1x_4^2+0.295745x_3x_4^2 \nonumber \\
&+0.153070x_4^3+0.354563x_1^2x_2^2+0.314975x_1x_2x_3x_4+0.274714x_1x_3^2x_4+0.316501x_2^4 \nonumber \\
&+0.358124x_3^4+0.311256x_1^4x_2+0.308158x_1^4x_3+0.394877x_1^3x_2^2+0.337582x_1^3x_4^2 \nonumber \\
&+0.325102x_1^2x_2x_3^2+0.317063x_1x_2^4+0.307461x_1x_2^2x_3^2+0.205988x_1x_2x_4^3-0.317399x_1x_4^4 \nonumber \\
&+0.369565x_2^2x_3x_4^2+0.366334x_2^2x_4^3+0.433464x_2x_4^4+0.363067x_3^5+0.299829x_1^6 \nonumber \\
&+0.308402x_1^5x_2+0.555666x_1^4x_4^2+0.359027x_1^3x_2^3+0.317469x_1^3x_2x_3^2+0.345314x_1^3x_3x_4^2 \nonumber \\
&+0.269574x_1^2x_4^4+0.425393x_3x_4^5+0.383478x_4^6+0.344957x_1^6x_3+0.294260x_1^6x_4 \nonumber \\
&+0.410708x_1^5x_3^2+0.285166x_1^3x_2x_3^3+0.190135x_1^3x_2x_4^3+0.099639x_1^3x_4^4+0.355064x_1^2x_2^4x_3 \nonumber \\
&+0.538182x_1^2x_4^5+0.266634x_1x_2^6+0.355968x_1x_2^4x_3^2+0.342439x_1x_2^3x_3^3+0.529434x_1x_2^2x_3x_4^3 \nonumber \\
&+0.505885x_1x_2x_3x_4^4+0.287929x_1x_3^6+0.423652x_1x_3^4x_4^2+0.580412x_1x_4^6+0.357337x_2^7 \nonumber \\
&+0.316656x_2^3x_4^4+0.305781x_2^2x_3^3x_4^2+0.248625x_2^2x_4^5+0.213728x_2x_3^3x_4^3
\label{eq:mackey23_2000_order7model}
\end{align}
where $x_1=x(t),~x_2=x(t-33),~x_3=x(t-15)$ and $x_4=x(t-45)$. 
\end{widetext}

\section{\label{sec:mackey80model}High-dimensional Mackey-Glass model}
\begin{widetext}
The $7^{th}$ degree polynomial model for Mackey-Glass system with parameters value $\gamma=0.1,~\beta=0.2,~n=10$ and $\tau=80$:
\begin{align}
x(t+1) = & 0.244052+1.276855x_1+0.203917x_3+0.204491x_7+0.274229x_9+0.216613x_1x_5 \nonumber \\
&+0.213592x_1x_7+0.355672x_1x_9+0.207846x_2x_7+0.210803x_3x_4+0.210518x_3x_6 \nonumber \\
&+0.202627x_3x_7+0.214811x_1x_3x_9+0.211629x_1x_4x_7+0.219037x_1x_5x_9+0.210260x_2x_8^2 \nonumber \\
&+0.228869x_3^2x_9+0.206613x_3x_6x_7+0.207033x_4x_5^2+0.210309x_5^2x_6+0.196956x_7^2x_9 \nonumber \\
&+0.186008x_7x_9^2+0.088274x_9^3+0.210147x_1^4+0.219800x_1x_2x_3x_9+0.219798x_1x_6x_7^2 \nonumber \\
&+0.211645x_2x_3^2x_9+0.206403x_2x_3x_6x_7+0.209316x_2x_7x_8^2+0.209678x_3x_5x_6x_7+0.211239x_4^2x_5x_9 \nonumber \\
&+0.208948x_5x_7x_8^2+0.130804x_9^4+0.208537x_1^4x_2+0.212468x_1^2x_6x_7^2+0.234417x_1x_2x_4x_5x_9 \nonumber \\
&+0.104559x_1x_9^4+0.217325x_4^3x_5x_9+0.212823x_1^2x_2^2x_7^2+0.208460x_1^2x_9^4+0.210994x_1x_3^2x_7x_9^2 \nonumber \\
&+0.210161x_1x_6x_9^4+0.206196x_4x_6x_9^4+0.207970x_1^4x_2x_4^2+0.195677x_1^4x_2x_5x_6+0.236119x_1^4x_3x_7x_9 \nonumber \\
&+0.210153x_1^4x_5^2x_6+0.204449x_1^3x_2^2x_4^2+0.175091x_1^3x_5x_9^3+0.216965x_1^2x_2^2x_7^3+0.201333x_1^2x_2x_9^4 \nonumber \\
&+0.205428x_1^2x_4x_6^3x_9+0.208188x_1^2x_5x_9^4+0.200196x_1^2x_8x_9^4+0.141584x_1^2x_9^5+0.214995x_1x_2^4x_5^2 \nonumber \\
&+0.212195x_1x_2^3x_5x_6x_8+0.213231x_1x_2^2x_5^2x_6x_8+0.216440x_1x_4^2x_6x_7^2x_8+0.211312x_1x_4x_5^2x_6x_8^2+0.210431x_1x_4x_6x_9^4 \nonumber \\
&+0.211375x_1x_5^6+0.213191x_1x_5^2x_6x_8^3+0.202278x_2^2x_3x_4x_6^2x_7+0.213338x_2x_4^4x_6^2+0.206811x_2x_4x_6^3x_7x_9 \nonumber \\
&+0.210530x_2x_6^6+0.212608x_2x_6^2x_7^4+0.204964x_3^7+0.210713x_3^6x_5+0.210644x_3^4x_5^2x_6 \nonumber \\
&+0.216775x_3^3x_6^3x_9+0.204288x_3^2x_4^2x_5^3+0.207530x_3x_4^6+0.217416x_3x_5x_6^4x_9+0.209786x_3x_7^6 \nonumber \\
&+0.256148x_3x_7^2x_9^4+0.211691x_4^4x_6^3+0.216558x_4^3x_6x_7^2x_8+0.206766x_4x_6^3x_7^2x_9+0.209552x_4x_7^5x_9 \nonumber \\
&+0.212761x_4x_8^6+0.210737x_5^2x_8^2x_9^3+0.210536x_5x_7x_8^2x_9^3+0.207928x_5x_8^2x_9^4+0.207856x_6x_8^2x_9^4 \nonumber \\
&+0.207171x_7^6x_8+0.209865x_7^6x_9+0.250657x_7x_9^6+0.313645x_9^7
\label{eq:mackey80model}
\end{align}
where $x_1=x(t),~x_2=x(t-63),~x_3=x(t-141),~x_4=x(t-106),~x_5=x(t-39),~x_6=x(t-85),~x_7=x(t-19),~x_8=x(t-125)$ and $x_9=x(t-157)$.
\end{widetext}

\begin{acknowledgments}
This research was supported by the Australian Research Council and Sirca Technology Pty Ltd under Linkage Project LP100100312.
The author is also supported by International Macquarie University Research Excellence Scholarship (iMQRES).
Deb Kane is thanked for her editorial support during preparation of this manuscript.
\end{acknowledgments}

\bibliography{pre4}

\end{document}